\documentclass[12pt]{article}
\usepackage[margin=1in]{geometry}
\usepackage{amsmath}
\usepackage{graphicx}
\usepackage{cite}
\usepackage{hyperref}
\hypersetup{colorlinks=true, linkcolor=blue, citecolor=blue, urlcolor=blue}
\usepackage{authblk}
\usepackage{tikz}
\usetikzlibrary{arrows.meta,positioning,fit,shapes,calc}
\usepackage{pgfplots}
\pgfplotsset{compat=1.18}
\usepgfplotslibrary{groupplots}
\usepackage{xcolor}

\title{Implicit and explicit treatments of model error in numerical simulation}

\author[1]{Danny Smyl\thanks{Corresponding author: \textit{danny.smyl@ce.gatech.edu}}}
\affil[1]{\small School of Civil and Environmental Engineering

Georgia Institute of Technology

Atlanta, GA, USA}
\date{}

\begin{document}
\maketitle

\begin{abstract}
Numerical simulations of physical systems  {exhibit} discrepancies  {arising from} unmodeled physics and idealizations {, as well as numerical approximation errors stemming from discretization and solver tolerances}. 
This article reviews techniques developed in the past  {several} decades to approximate and account for model errors, both implicitly and explicitly. 
Beginning from fundamentals, we frame model error in inverse problems, data assimilation, and predictive modeling contexts. 
We then survey major approaches: the Bayesian approximation error framework, embedded internal error models for structural uncertainty, probabilistic numerical methods for discretization uncertainty, model discrepancy modeling in Bayesian calibration and its recent extensions, machine-learning-based discrepancy correction, multi-fidelity and hybrid modeling strategies, as well as residual-based, variational, and adjoint-driven error estimators. 
Throughout, we emphasize the conceptual underpinnings of implicit versus explicit error treatment and highlight how these methods improve predictive performance and uncertainty quantification in practical applications ranging from engineering design to Earth-system science. Each section provides an overview of key developments with an extensive list of references to facilitate further reading. 
The review is written for practitioners of large-scale computational physics and engineering simulation, emphasizing how these methods can be incorporated into PDE solvers, inverse problem workflows, and data assimilation systems.
\end{abstract}

\pagebreak

\section{Introduction}
\begin{figure}[h!]
\centering
\begin{tikzpicture}

\begin{axis}[
  name=plotA,
  width=0.675\linewidth,
  height=5.0cm,
  xmin=0, xmax=10,
  ymin=-1.2, ymax=1.2,
  grid=both,
  major grid style={gray!40},
  minor grid style={gray!20},
  minor tick num=1,
  ticklabel style={font=\small},
  label style={font=\small},
  title style={font=\small},
  ylabel={Response},
  title={(a) True model vs.\ numerical model},
  legend style={
    font=\small,
    cells={anchor=west},
    draw=none,
    fill=none,
    at={(0.01,0.01)},
    anchor=south west
  }
]

\addplot[
  blue,
  very thick,
  domain=0:10,
  samples=300
] {sin(deg(x))};
\addlegendentry{True model}

\addplot[
  black,
  dashed,
  thick,
  domain=0:10,
  samples=300
] {sin(deg(x)) + 0.15*cos(deg(0.5*x))};
\addlegendentry{Numerical model}

\end{axis}

\begin{axis}[
  at={(plotA.south)},
  anchor=north,
  yshift=-1.8cm,
  width=0.675\linewidth,
  height=5.0cm,
  xmin=0, xmax=10,
  ymin=-1.2, ymax=1.2,  
  grid=both,
  major grid style={gray!40},
  minor grid style={gray!20},
  minor tick num=1,
  ticklabel style={font=\small},
  label style={font=\small},
  title style={font=\small},
  xlabel={$x$},
  ylabel={Model error},
  title={(b) Model error (true model $-$ numerical model)}
]

\addplot[
  red!70!black,
  thick,
  domain=0:10,
  samples=300
] { sin(deg(x)) - (sin(deg(x)) + 0.15*cos(deg(0.5*x))) };

\addplot[
  gray!60,
  thin,
  domain=0:10,
  samples=2
] {0};

\end{axis}

\end{tikzpicture}
\caption{Illustrative one-dimensional numerical model error schematic.
Panel~(a) shows a numerical model and corresponding true model. Panel~(b) shows the
 model error on the same vertical scale,
highlighting that the numerical model error is small and structured -- not random.}
\label{fig:true-vs-model-error}
\end{figure}
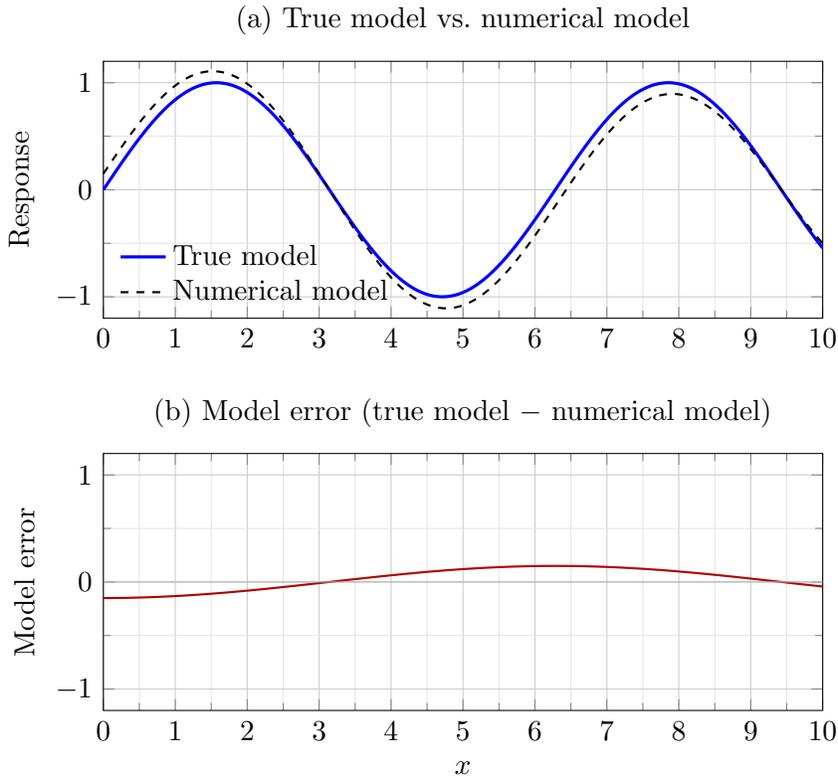

All computational models are imperfect abstractions of reality. As the famous adage goes, \emph{``all models are wrong, but some are useful''} -- the gap between a model's predictions and the true system is broadly termed \emph{model error} or \emph{model discrepancy}. 
Model errors, illustrated schematically in Fig. \ref{fig:true-vs-model-error}, can arise from incomplete physics in the model (model-form errors), and numerical approximation (discretization errors) \cite{Roy2011, Oberkampf2002};  {additional contributions to total prediction error come from uncertain or misspecified parameters.}
Herein,  {we review major sources of discrepancy} and distinguish two key strategies for handling model error:
\begin{itemize}
  \item \textbf{Implicit methods}, which treat model error statistically (e.g.\ inflating uncertainty to absorb inaccuracies) without directly modifying the model equations.
  \item \textbf{Explicit methods}, which introduce additional error correction terms or models (learned or analytic) to represent the discrepancy.
\end{itemize}

We stress that in this article “implicit” and “explicit” refer to how model error is represented in the statistical or computational formulation, not to implicit/explicit time-integration schemes. 
By “implicit” we mean that model error is absorbed into uncertainty descriptions (e.g. inflated noise or process covariances) without altering the governing equations, whereas “explicit” refers to introducing additional variables, operators, or correction terms that represent the discrepancy itself. 
Some methods sit in between these extremes; wherever classification is ambiguous, we group them according to how they are typically used in practice.

In contrast to earlier reviews that focus primarily on verification and validation frameworks or on specific model classes, the present article offers a unifying perspective on model error across several communities. 
Our main contributions are: (i) to organize model error treatments systematically into implicit and explicit strategies and relate this taxonomy to concrete numerical workflows; (ii) to synthesize developments spanning Bayesian inverse problems, data assimilation, probabilistic numerics, multi-fidelity modeling, and machine-learning-based correctors; and (iii) to highlight how these methods can be integrated into large-scale computational physics codes for predictive simulation and uncertainty quantification.

We begin by clarifying the types of model error and their treatment. Sections~\ref{sec:implicit} and~\ref{sec:explicit} then survey implicit and explicit error approximation techniques, respectively. In each category, we discuss foundational methods and subsequent advancements, including Bayesian approximation error modeling, embedded error representations, probabilistic numerical approaches, statistical calibration with discrepancy, machine-learning-based correctors, multi-fidelity models that explicitly model the gap between fidelity levels, and methods that use residuals or adjoint equations to estimate error. Section~\ref{sec:applications} touches on applications of these ideas in data assimilation (DA), inverse problems, and PDE-constrained optimization, highlighting how accounting for model error improves inference and design \cite{Carrassi2018, Allaire2014}. We conclude with perspectives on current challenges and future directions.

\section{ {Definitions and Error Decomposition}}\label{sec:fundamentals}
\begin{figure*}[h!]
\centering
 \includegraphics[width=15cm]{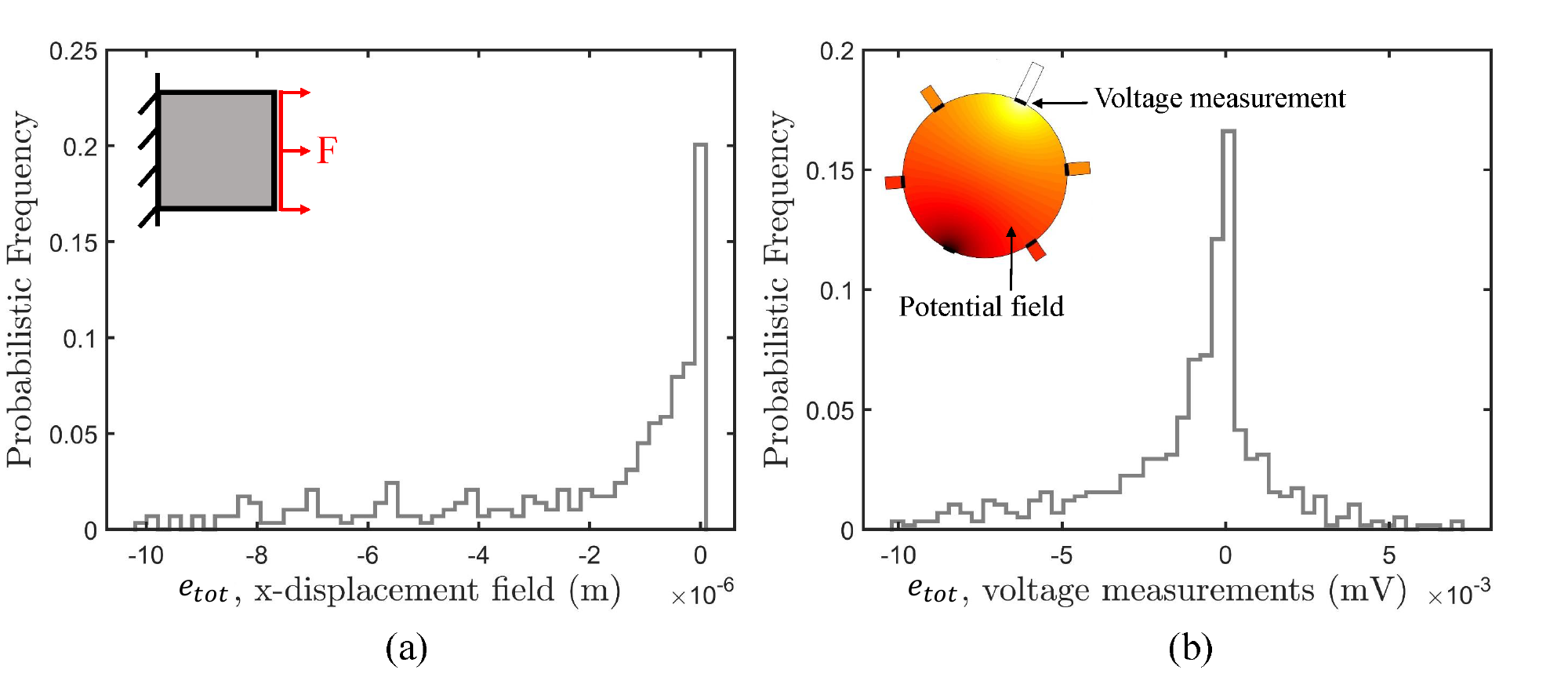} 
\hfill
\caption{Error histograms from \cite{Smyl2021} showing the  error frequencies for (a) the displacement field of a clamped steel plate deformed in the horizontal direction and (b) electrostatic boundary voltage simulations from an inhomogeneous conductivity circular domain. In both cases, error distributions are markedly non-Gaussian and not well captured by Gaussian models.}
\label{hist}
\end{figure*}

It is important to delineate what we mean by \emph{model error}. In broad terms, model error  {(sometimes called model inadequacy or structural uncertainty when referring to model-form error)} is the discrepancy between the true physical process and the mathematical/computational model used to represent it \cite{Roy2011,mueller2012linear}. This is separate from uncertainties in model \emph{inputs} (such as parameter uncertainty) and from measurement or observation errors, $\epsilon$ \cite{Iakovidis2021,Kaipio2007,Stuart2010}.
Moreover, unlike measurement error which is often  {idealized} as Gaussian, model error is  {frequently structured and can be markedly} non-Gaussian as shown in Fig. \ref{hist}.

A  {classical} way to formalize the basic error decomposition is
\begin{equation}
  y^{\text{true}}(x)
  \;=\;
  f(x,\theta^\star) + \delta(x),
  \label{eq:basic-decomp}
\end{equation}

\noindent where $y^{\text{true}}(x)$ is the true system response at input $x$, $f(x,\theta^\star)$ is the output of the  {idealized continuum physics (or exact mathematical operator)} evaluated at the ``true'' parameters $\theta^\star$, and $\delta(x)$ is the \emph{model discrepancy}\footnote{ {We note that $\delta(x)$ may also be parameterized with dependence on $\theta$}.} capturing structural/model-form error.
 {One can, however, heuristically} separate the total prediction error into  {individual} components  {for a real discretized numerical model $y_h$},

\begin{equation}
  e_{\text{tot}}
  = y^{\text{true}}(x) - y_h\bigl(x,{\theta}\bigr)
  \approx  e_{\text{struct}} + e_{\text{param}} + e_{\text{num}},
  \label{eq:error-split}
\end{equation}

\noindent  where \(y_h\) is evaluated at the inferred (or calibrated) parameter vector \(\theta\). The term \(e_{\text{struct}}\) encodes model-form inadequacy (missing physics, wrong constitutive laws), \(e_{\text{param}}\) arises from parameter misspecification (\(\theta\neq\theta^\star\)), and \(e_{\text{num}}\) from numerical discretization and solver error.
We emphasize that this decomposition is not unique in practice: structural, parametric, and numerical effects can be partially confounded, and only certain combinations are statistically identifiable from data \cite{Brynjarsdottir2014,Kennedy2001}.

Sources of model error include, for example:
\begin{itemize}
    \item \textbf{Unmodeled or approximated physics:} e.g.\ turbulence closures, constitutive laws, or boundary conditions that do not perfectly represent reality \cite{Duraisamy2019, Cheung2011}.
    \item \textbf{Finite resolution and discretization:} numerical grid or time-step coarseness leading to unresolved subscale effects and discretization error \cite{Roy2011,surana2017finite,SuranaJoy2016}.
    \item \textbf{Unknown boundary or initial conditions:} errors due to misspecified domain geometry or initial states \cite{Kaipio2007,mueller2012linear}.
    \item \textbf{Numerical approximations:} algorithmic errors from truncation, linearization, or solver convergence tolerances \cite{Oberkampf2002,surana2016finite}.
\end{itemize}
For instance, in weather prediction and climate models, unresolved sub-grid processes (like convection) and their parametrizations are major sources of model error \cite{Carrassi2018,ParametrizationinWeather}. In engineering simulations, using a reduced-order or surrogate model introduces bias relative to high-fidelity physics; quantifying and correcting such reduced-model errors is a central theme in reduced-order modeling (ROM) and multifidelity uncertainty quantification (UQ) \cite{Drohmann2015, Peherstorfer2018}.

If untreated, model error can manifest as systematic bias or as state- and time-dependent deviations that lead to overconfident and biased predictions. 
This has been emphasized in verification, validation, and UQ frameworks \cite{Roy2011, Oden2013, Oberkampf2002} and in statistical calibration of computer models \cite{Kennedy2001, Brynjarsdottir2014}. Brynjarsd\'ottir and O'Hagan \cite{Brynjarsdottir2014} note that ignoring model discrepancy when calibrating models to data can severely mislead inference of physical parameters. Inverse problems are notoriously sensitive to unmodeled errors, often leading to misfitting if model inadequacy is not accounted for \cite{Kaipio2007, Stuart2010}.

Approaches to model error can be classified as implicit or explicit. 
Written plainly, in an implicit approach, one does not attempt to explicitly build a term for model error; instead, one inflates or adjusts uncertainty in parameters or observations so that the effects of model error are statistically accommodated. A classic example is inflating the observation noise variance or introducing process noise to account for unmodeled effects \cite{Bayarri2007, Mitchell2015, Carrassi2018}. By contrast, {explicit approaches} introduce additional structures or variables to represent the model error itself. 
For example, one might add a discrepancy function $\tilde\delta(x)$ so that  { for a measurement $z$ taken at input $x$, the observation model becomes}

\begin{equation}
  { z \;=\; y_h(x,\theta)} + \tilde\delta(x) + \epsilon,
  \label{eq:discrepancy}
\end{equation}

\noindent with $\tilde\delta(x)$, to be determined or learned \cite{Kennedy2001, Higdon2008},  { representing (in principle) the total discrepancy, encompassing both structural inadequacy and the numerical discretization error of $y_h$.}
Machine-learning approaches that learn a corrective mapping from inputs (and possibly states) to model error are explicit \cite{Duraisamy2019, Smyl2021,lunz2021learned}. The choice of implicit vs.\ explicit treatment often depends on context: explicit methods can provide deeper insight and better corrections but at the cost of additional modeling complexity, whereas implicit methods are convenient when an adequate stochastic description of unresolved errors is available.


Throughout the review we concentrate on settings typical of computational physics and engineering: large-scale numerical solution of PDE and ODE models, often coupled to inverse problems, data assimilation schemes, or optimization loops. 
Our examples and references are drawn primarily from applications such as fluid and solid mechanics, geophysics, climate and weather modeling, and related fields.

 {\textbf{Remark:} In this review, a \emph{model} refers broadly to a predictive map from inputs (and/or states) to quantities of interest. 
This includes physics-based models derived from governing laws and implemented numerically, as well as data-driven models learned from data (including purely statistical/learned predictors and hybrid physics-learned formulations). 
The dominant sources of model error differ across these settings, but the implicit/explicit distinction reviewed here applies to both.}

\section{Implicit Approaches to Model Error Approximation}\label{sec:implicit}
Implicit methods account for model error by adjusting the probabilistic framework of the problem, rather than by modifying the model's functional form. These methods treat the effect of model inadequacy as an additional uncertainty to be quantified. We discuss three major implicit approaches: the Bayesian approximation error technique, probabilistic numerical methods, and methods in data assimilation that treat model error as noise.

\subsection{Bayesian Approximation Error (BAE) Technique}
One influential implicit approach is the \emph{Bayesian approximation error} (BAE) framework developed by Kaipio, Somersalo, and collaborators \cite{Kaipio2007, Huttunen2007, kolehmainen2009approximation,lipponen2018correction,nissinen2010compensation}. This approach acknowledges that we often use simplified forward models in inverse problems (for computational tractability  {and speed}) and thus introduces a discrepancy between the simplified model and the ``true'' physics.

Let $y_H(\theta)$ and $y_L(\theta)$ denote the high-fidelity and approximate model outputs, respectively. The approximation (or \emph{model}) error is
\begin{equation}
  \tilde{e}(\theta) \;=\; y_H(\theta) - y_L(\theta).
  \label{eq:bae-error}
\end{equation}
In the BAE formulation, $\tilde{e}(\theta)$ is treated as a random quantity with mean $\mu_e$ and covariance $\Gamma_e$ estimated from offline paired simulations or analytical arguments. Suppose the original observation model with measurements $z$ would be

\begin{equation}
  z = y_H(\theta) + \epsilon, \qquad \epsilon \sim \mathcal{N}(0,\Gamma_\epsilon).
\end{equation}

\noindent Using the computationally cheaper model and approximation error gives

\begin{equation}
  z = y_L(\theta) + \tilde{e}(\theta) + \epsilon \;\approx\; y_L(\theta) + \eta,
  \qquad
  \eta \sim \mathcal{N}(\mu_e,\, \Gamma_\epsilon + \Gamma_e),
  \label{eq:bae-like}
\end{equation}

\noindent i.e.\ the approximation error is absorbed into an augmented term with mean $\mu_e$ and covariance $\Gamma_\epsilon+\Gamma_e$. 
Under the Gaussian closure in Eq.~\eqref{eq:bae-like}, premarginalizing over $\tilde e$ yields a modified likelihood with shifted mean and inflated covariance, often leading to a more realistic posterior for $\theta$ \cite{Kaipio2007, Huttunen2007}. 
 {In general, it is important to note that $\tilde e(\theta)$ can be $\theta$-dependent and correlated with $y_L(\theta)$ (and, since $y_H(\theta)=y_L(\theta)+\tilde e(\theta)$, also with $y_H(\theta)$); with explicit inputs, the same holds for $\tilde e(x,\theta)$ with $y_L(x,\theta)$ and $y_H(x,\theta)$. In such a case where this may need accounting for, additional statistical considerations will need to be made.}

The BAE approach has been successful, having been applied to a wide range of inverse problems. Initial applications were in electrical impedance tomography (EIT) and optical tomography, where model reduction (using coarse meshes or simplified physics) is common. 
Arridge \emph{et al.}\ \cite{Arridge2006} and Huttunen, Heikkinen, Nissinen, Tarvainen, Vauhkonen, Kolehmainen  \& Kaipio \cite{Huttunen2007,nissinen2010compensation,tarvainen2009approximation,nissinen2007bayesian} showed that compensating for modeling errors via BAE reduced biases in reconstructions, allowing the use of coarse grids or truncated domains while maintaining accuracy. Related work handles high-dimensional uncertainties and sparse priors \cite{Koutsourelakis2009, Marzouk2007}. Approximation error ideas have also been exploited in multi-fidelity and real-time inversion settings \cite{Absi2016, Biehler2015}.

A practical limitation of BAE is the need to characterize the distribution of $\tilde{e}$. This is often done via offline simulation: one runs the high- and low-fidelity models on a training set of inputs to collect $\tilde{e}$ samples and then estimates $\mu_e$ and $\Gamma_e$ \cite{Huttunen2007, Marzouk2007}. 
In complex problems, this can be costly, and if the model discrepancy is highly non-Gaussian or input-dependent, a simple Gaussian model may be inadequate. More flexible, possibly hierarchical formulations have been proposed to relax Gaussianity and to update error statistics adaptively \cite{Absi2016}. Nevertheless, BAE provides a powerful, conceptually straightforward way to implicitly handle model errors in Bayesian inverse problems and has been adopted in biomedical imaging, process engineering, and structural health monitoring \cite{Kaipio2007, Arridge2006, Absi2016, Biehler2015}.

 {It should be remarked that many of the same ideas in this subsection extend naturally to data-driven models: learned surrogates can be used as the low-fidelity mapping in BAE, with approximation-error statistics correcting systematic surrogate bias. 
Likewise, for learned continuous-time dynamical models including NeuralODEs \cite{ko2023homotopy,yildiz2019ode2vae,chen2018neural}, discretization and solver tolerances introduce numerical approximation error that can be treated using similar probabilistic/numerical-error considerations discussed for physics-based solvers.}

\subsection{Probabilistic Numerics and Uncertainty from Discretization}
Classical numerical algorithms (for integration, differential equations, linear algebra) produce a single deterministic output, often ignoring the fact that discretization and finite termination introduce error. \emph{Probabilistic numerics} is an emerging paradigm that instead models numerical uncertainty as part of the computation \cite{Hennig2015, Cockayne2019}. 
The basic idea is to design numerical methods that return not just an estimate but also a probability distribution over the true quantity, quantifying numerical error. This can be seen as an implicit model error approach: the numerical approximation error is treated as a random component, which can be propagated through subsequent computations or inference tasks.

For example, consider solving an ODE
\begin{equation}
  \frac{du}{dt} = g(u,t), \qquad u(0) = u_0,
  \label{eq:ode}
\end{equation}

\noindent or computing an integral $I = \int f(x)\,dx$. 
Probabilistic ODE solvers and Bayesian quadrature methods place a prior (often a Gaussian process) on the unknown solution or integrand, then update this prior using residuals or function evaluations, yielding a posterior distribution for $u(t)$ or $I$ \cite{Hennig2015, Cockayne2019}. 
Early work by O'Hagan in the 1990s framed numerical integration (quadrature) as a Bayesian inference problem, treating the unknown integrand as a random function and updating a prior with point evaluations to obtain a posterior over the integral itself \cite{OHagan1992}. This perspective has been generalized by subsequent authors to a broad class of numerical tasks. More recently, Hennig et al.\ \cite{Hennig2015} advocated viewing computation itself as inference: for instance,  {for symmetric positive definite linear systems (under an appropriate Gaussian prior/observation model)}, one can solve linear systems by iteratively updating a Gaussian belief over the solution vector such that the mean follows the conjugate gradient algorithm and the covariance reflects uncertainty after a finite number of iterations. In a similar spirit, probabilistic ODE solvers use Gaussian process priors for the unknown solution function and update them with information from the ODE residuals \cite{Hennig2015, Cockayne2019}. The result is a mean trajectory (the conventional numerical solution) and a credibility band representing discretization error.

Conrad \emph{et al.}\ \cite{Conrad2017} formalized the notion of probability measures for numerical solutions of differential equations, quantifying epistemic uncertainty induced by discretization and showing how this uncertainty can be propagated into subsequent inference. Such methods yield a mean trajectory (the conventional numerical solution) together with a credibility band representing discretization error. This uncertainty can be used to adaptively refine computations (e.g.\ refining the grid until the variance falls below a tolerance) and to account for numerical error when solving inverse problems with expensive forward solvers \cite{Cockayne2019}. 

In summary, probabilistic numerics implicitly handles one important source of model error -- numerical solution error -- by treating computation as inference and providing a quantified measure of trust in the results. When coupled with other UQ components, these methods help ensure that downstream analyses (e.g.\ optimization or inference) are not blindly trusting a possibly inaccurate computation \cite{Hennig2015, Cockayne2019, Oberkampf2002}.

\subsection{Model Error as Noise in Data Assimilation}\label{sec:DA}
Data assimilation (DA) is a field where model error has long been recognized and treated -- typically implicitly -- by inflating process noise or using weak constraints. 
In sequential state estimation (such as the Kalman filter and its variants), one commonly assumes a state evolution model

\begin{equation}
  x_{k+1} = M(x_k) + \xi_k,
  \qquad \xi_k \sim \mathcal{N}(0,Q),
  \label{eq:da-model}
\end{equation}

\noindent together with an observation model

\begin{equation}
   {z_k} = H(x_k) + \epsilon_k,
  \qquad \epsilon_k \sim \mathcal{N}(0,R).
  \label{eq:da-obs}
\end{equation}

\noindent Here \(M(\cdot)\) is the forecast/state-transition operator advancing  {\(x_k \mapsto x_{k+1}\), \(H(\cdot)\) is the observation operator mapping the state to observation space, \(\xi_k\) is the process (model) ``noise", and $R$ is the observation-error covariance.}
If the physical model were perfect, one might set \(\xi_k = 0\);
 however, to account for model imperfections, DA practitioners introduce a nonzero process noise covariance $Q$, effectively treating model error as a random forcing term \cite{Mitchell2015, Carrassi2018}. The choice of $Q$ is crucial: if $Q$ is too small, the filter/smoother can become overconfident and misfit the data; if $Q$ is too large, one may attribute too much discrepancy to model noise and underutilize the dynamics.

Another class of DA methods, \emph{weak-constraint 4D-Var}, addresses model error in a variational setting. In strong-constraint 4D-Var, one assumes the model trajectory is exact (no model error), and only initial conditions are adjusted to fit observations. 
Weak-constraint 4D-Var relaxes this assumption by allowing a model error term at each time step, which is estimated alongside the initial state \cite{Tremolet2006}. The 4D-Var cost function includes a term of the form

\begin{equation}
  \mathcal{C}_{\text{model}}(\{\xi_k\}) = \frac{1}{2}\sum_k \xi_k^\top Q^{-1}\xi_k,
\end{equation}

penalizing large model corrections but not forbidding them. 
This formulation has been implemented in research/experimental settings to account for systematic model biases \cite{Tremolet2006, Carrassi2018}.

In ensemble-based DA methods, related practices include \emph{covariance inflation} and \emph{stochastic model perturbations}, where one artificially inflates ensemble spread or adds random tendencies as pragmatic proxies for under-dispersion arising from multiple sources (including model error and finite-ensemble/sampling effects) \cite{Mitchell2015, Carrassi2018}.
More recently, machine-learning-assisted DA approaches have emerged: Farchi \emph{et al.}\ \cite{Farchi2021}  {combined ML with DA to learn model-error correction from sparse/noisy observations, yielding a hybrid model that improves short- to mid-range forecasts and the resulting DA analyses.}

All these DA techniques treat model error implicitly by broadening the uncertainty in the forecasting step: the dynamical model $M$ is not fundamentally changed; instead, additional variance or random perturbations are introduced so that the assimilation algorithm recognizes that the forecast cannot be trusted fully. This has been useful in chaotic systems like the atmosphere, where even small unmodeled processes can lead to large forecast deviations \cite{Carrassi2018, Mitchell2015}. 
A major ongoing challenge is specifying or estimating the statistics of model error ($Q$ and potential bias terms) from data in a stable and robust way; innovation-based methods, hierarchical models, and learned parameterizations are active research directions \cite{Carrassi2018, Farchi2021}.

\section{Explicit Approaches to Model Error Approximation}\label{sec:explicit}
Explicit methods introduce additional corrective structures into the modeling framework to represent model error. Rather than solely relying on adjusted uncertainties, these approaches attempt to learn, calibrate, and/or otherwise determine a \emph{model error term} that, when added to the original model, yields improved agreement with reality.

\subsection{Model Discrepancy in Bayesian Calibration (Kennedy-O'Hagan Framework)}
Perhaps the most widely cited explicit treatment of model error is the framework by Kennedy and O'Hagan (KOH) \cite{Kennedy2001}. In their seminal work on Bayesian calibration of computer models, they introduced an additive discrepancy function, recalling Eq.~\eqref{eq:discrepancy}, 
where, for explicit errors: $y_h(x,\theta)$ is the computer model prediction given input $x$ and calibration parameters $\theta$, $\epsilon$ is measurement noise, and $\tilde\delta(x)$ is the model discrepancy. 
The discrepancy term $\tilde\delta(x)$ absorbs systematic differences between the model (even at the best-fit parameters) and the true system.

Kennedy and O'Hagan \cite{Kennedy2001} placed a Gaussian process (GP) prior on $\tilde\delta(x)$ to encode beliefs about smoothness and variability. Calibration then jointly infers $\theta$ and $\tilde\delta(x)$ from data. 
This framework distinguishes parametric uncertainty from model-form uncertainty, which is important because without $\tilde\delta(x)$ the calibration of $\theta$ will attempt to force the model to fit the data even if the model structure is wrong, often resulting in biased parameter estimates \cite{Brynjarsdottir2014}.

However, this flexibility introduces \emph{non-identifiability}: if $\tilde\delta(x)$ is too flexible, it can soak up variation that might otherwise inform $\theta$ \cite{Brynjarsdottir2014, Tuo2015, Plumlee2017}. Strategies to mitigate confounding include:
\begin{itemize}
  \item Using multiple outputs or multiple operating conditions (multi-response calibration) to better distinguish parameter effects from discrepancy \cite{Arendt2012, Higdon2008}.
  \item Structuring the prior for $\tilde\delta(x)$ (e.g.\ smoothness, low-dimensional basis) to capture broad trends but avoid overfitting \cite{Kennedy2001, Plumlee2017}.
  \item Focusing on predictive calibration rather than strict parameter identifiability, as argued by Plumlee \cite{Plumlee2017}.
\end{itemize}
The KOH framework underpins many validation and uncertainty quantification methodologies. Bayarri \emph{et al.}\ \cite{Bayarri2007} built a validation framework using KOH-type models, and Higdon \emph{et al.}\ \cite{Higdon2008} demonstrated high-dimensional calibration with principal component representations of outputs. Extensions have incorporated structured discrepancy models informed by physics, such as spatially varying corrections to RANS Reynolds stresses in turbulence modeling \cite{Xiao2016, Zhang2018}.

\subsection{Gaussian Process Surrogates and Emulators for Model Error}
Gaussian processes not only serve as priors for discrepancy functions in calibration, but also underpin many surrogate modeling techniques where one explicitly emulates the difference between a high-fidelity and low-fidelity model. In multi-fidelity modeling, one often has a hierarchy of models $\{f_0, f_1, \dots, f_H\}$ of increasing fidelity and cost. A popular approach is GP-based co-kriging, where one writes the high-fidelity model as

\begin{equation}
  f_H(x) = \rho\, f_L(x) + \delta_{\text{mf}}(x),
  \label{eq:mf}
\end{equation}

\noindent with $f_L(x)$ a low-fidelity model, $\rho$ a scale factor, and $\delta_{\text{mf}}(x)$ a GP representing the difference (model error) between low and high fidelity \cite{OHagan2000, LeGratiet2013, Peherstorfer2018}. 
 {Here, $f_L(x)$ and $f_H(x)$ are similar to $y_L(x)$ and $y_H(x)$, but are delineated to emphasize that they represent low- and high-fidelity simulator outputs used for cross-model emulation, rather than the observation model outputs in the BAE setting.}
This is analogous to Eq.~\eqref{eq:discrepancy} but now the discrepancy is between fidelities rather than between model and reality.

Such multi-fidelity GPs have been widely applied in aerospace design, structural optimization, and climate modeling \cite{Perdikaris2017, Peherstorfer2018}. Perdikaris \emph{et al.}\ \cite{Perdikaris2017} proposed nonlinear autoregressive GP models that iteratively map low- to high-fidelity outputs, effectively learning a series of discrepancy functions. Multi-fidelity Bayesian optimization and Monte Carlo methods similarly exploit these error models to reduce the cost of optimization or uncertainty propagation \cite{ghoreishi2016uncertainty,Peherstorfer2018, Nagel2016}.

GPs are also used as stand-alone surrogates for model error in situations where one compares a single computational model to experimental data. One can fit a GP to residuals $r_i = y_i - f(x_i)$ and then use this GP to correct future model outputs \cite{Kennedy2001, Rasmussen2006}. This amounts to learning $\tilde \delta(x)$ in Eq.~\eqref{eq:discrepancy} in a non-parametric fashion. Applications include bias correction of engineering simulations (e.g.\ CFD, engine models) and spatially distributed error fields in turbulence modeling \cite{Zhang2018, Cheung2011}.

One benefit of GP-based explicit error models is their flexibility and smoothness properties; they can capture complex error patterns with relatively few hyperparameters and provide uncertainty estimates for the error at new points, aligning with UQ goals \cite{Rasmussen2006, Peherstorfer2018}. Limitations include scalability and the assumption of smoothness; non-smooth or highly nonstationary errors may require tailored kernels, sparse approximations, or alternative models such as deep GPs.

\subsection{Embedded Error Models in Equations}
While the discrepancy approaches above add corrections at the output level, an alternative is to embed model error representations \emph{within the model equations}. The idea is to introduce extra degrees of freedom directly into the governing equations or source terms, representing unknown physics.

Sargsyan \emph{et al.}\ \cite{Sargsyan2019} provide a clear example of this strategy. They consider computational models of combustion chemistry where certain reaction pathways are not well characterized, leading to  model error. 
Instead of treating this as an output bias, they augment kinetics with correction terms parameterized by polynomial chaos expansions,  {which we write schematically using an additive term},

\begin{equation}
   {\dot{q}} = F( {{q}},\theta) + \sum_{j} \alpha_j \psi_j(\xi),
\end{equation}

\noindent where  {$q$ denotes the state vector,} $\psi_j$ are the $j^{th}$ basis functions (e.g.\ orthogonal polynomials in latent variables $\xi$) and $\alpha_j$ are unknown coefficients inferred from data. This embedded representation is calibrated jointly with physical parameters in a Bayesian framework, disentangling parametric and structural uncertainty \cite{Sargsyan2019, Oden2013}.

Embedded approaches have been explored in climate modeling (adding correction terms in subgrid closure schemes and calibrating those), in groundwater flow (adding unknown source terms representing wells or leaks), and in CFD (embedding eddy-viscosity fields or forcing terms to represent unknown effects) \cite{Duraisamy2019, Oden2013}. A related concept in state-space models is \emph{state augmentation}, where one augments the state with additional variables representing model bias and prescribes dynamics for them, then infers these via filtering \cite{Jazwinski, Carrassi2018}. Embedded error representations can maintain physical consistency and interpretability and often provide insight that can guide future model development.

\subsection{Machine Learning for Model Error: Deep Learning and Hybrid Models}
In recent years, there has been  {growing} interest in using machine learning (ML), especially deep learning, to learn complex model error patterns from data. 
These approaches are explicit in that they directly learn a mapping to correct the model, but they often function as black boxes, using data to capture discrepancies that might be intractable to derive analytically.
 {Moreover, while many examples in this section treat ML as a corrector on top of a mechanistic core, we emphasize that data-driven models are also deployed as the primary predictive model; in that case, ``model error” can manifest as generalization error and structural misspecification, and the same implicit/explicit treatments can be interpreted as uncertainty absorption versus explicit discrepancy correction.}

One prominent class of methods is \emph{hybrid modeling} or \emph{physics-informed machine learning}, where a physics-based model is combined with an ML model that predicts corrections (a ``delta'' model). In turbulence modeling, for example, neural networks have been trained to predict discrepancies in Reynolds stresses or mean velocity profiles given flow features, effectively creating closure models that correct baseline RANS simulations \cite{Duraisamy2019, Xiao2016, Zhang2018}. 
 {In published case studies, these hybrid closures can reduce error in quantities of interest} such as lift and drag.
In recent engineering applications, Swischuk and Allaire \cite{swischuk2019machine} applied a machine-learning approach to detect and correct aircraft sensor errors, demonstrating how data-driven correction operators can explicitly compensate for measurement or model input anomalies. 

In geophysical modeling, Brajard \emph{et al.}\ and Farchi \emph{et al.}\ combined DA and ML to learn model error in chaotic dynamical systems \cite{Brajard2021, Farchi2021}. A typical cycle alternates between a DA step, which yields an analysis trajectory, and an ML step, which trains a neural network to map the model state to the local tendency error (difference between analysis tendency and model tendency). When deployed, the learned error model corrects the forecast tendencies, improving forecast skill and stability relative to the baseline model \cite{Brajard2021, Farchi2021}.

Deep learning has also been used to represent subgrid terms in climate models, e.g.\ using convolutional neural networks to emulate convection or cloud processes, and in engineering systems as part of \emph{digital twins}. Pathak \emph{et al.}\ \cite{Pathak2018} constructed a hybrid forecast model for the Kuramoto--Sivashinsky equation by coupling a partial physics model with a reservoir computing approach that learned unresolved dynamics, extending the forecast horizon beyond what the physics-only model could achieve. Willard \emph{et al.}\ \cite{Willard2022} survey many such approaches, emphasizing the importance of encoding known physics (e.g.\ invariances, conservation laws) in the ML architecture or loss to ensure physically consistent corrections.

Recent studies by Hauptmann and collaborators propose learned model-correction operators that explicitly compensate for operator approximation error within inverse problems and reconstruction workflows \cite{Hauptmann2023, ArjasHauptmann2023, lunz2021learned}. For example, the ‘model-corrected learned primal-dual’ framework embeds a correction network alongside an approximate forward model, yielding improved fidelity under limited data or coarse modeling assumptions \cite{Hauptmann2023}.

Challenges with ML-based discrepancy models include generalization beyond the training domain, data requirements, and maintaining physical constraints. Hybrid strategies that keep physics in the loop and task ML with learning only the discrepancy (rather than the full dynamics) appear effective and robust \cite{Duraisamy2019, Willard2022, Smyl2021}. Moreover, the recent rise of physics-informed neural networks (PINNs) -- which embed the governing PDE/ODE residuals directly into the neural network training loss \cite{raissi2019physics, qian2020lift, karniadakis2021physics, zhuang2025physics} -- offers a promising avenue for discrepancy modeling: by leveraging known physics to regularize learning, one can mitigate over-fitting to limited training data, improve generalizability into unseen regimes, and enforce constraint‐respecting corrections. 

\subsection{Multi-Fidelity and Hybrid Surrogate Modeling}
Multi-fidelity modeling refers to strategies that combine simulations or models of varying levels of detail and accuracy to achieve predictions with improved efficiency \cite{Peherstorfer2018, FernandezGodino2023}. In the context of model error, multi-fidelity approaches explicitly approximate the error of a low-fidelity model by leveraging a high-fidelity model.

The fundamental insight is that a low-fidelity model $f_L$ (e.g.\ coarse-mesh CFD or an analytical simplification) will have a discrepancy relative to a high-fidelity model $f_H$ that often varies more smoothly or simply than $f_H$ itself. By learning $\delta_{\text{mf}}(x)$ in Eq.~\eqref{eq:mf} from a limited number of high-fidelity runs, one can cheaply correct many low-fidelity predictions \cite{OHagan2000, LeGratiet2013, Perdikaris2017}. This idea underlies a broad class of multi-fidelity Monte Carlo, Bayesian optimization, and surrogate modeling techniques \cite{Peherstorfer2018, Nagel2016}.

Hybrid surrogate models can also combine data-driven and physics-driven components. For instance, Allaire and Willcox \cite{Allaire2014} integrated Bayesian calibration into a multi-fidelity aerospace design loop, calibrating a low-fidelity model to high-fidelity data (using discrepancy terms) and then performing optimization with the calibrated low-fidelity model, periodically validating with high-fidelity simulations.  Fernandez-Godino \cite{FernandezGodino2023} provides a comprehensive review of multi-fidelity models, emphasizing the treatment of cross-fidelity error.

The success of multi-fidelity methods depends on strong correlation between fidelities. When correlation is weak or qualitatively different behavior occurs (e.g.\ missing modes), multi-fidelity corrections may be insufficient, and intermediate fidelities or more expressive discrepancy models (e.g.\ deep operator networks) may be necessary \cite{Ahmed2023, Peherstorfer2018}. Nonetheless, when applicable, these approaches can turn model error into a resource for variance reduction and computational savings.

\subsection{Residual-Based and Adjoint-Based Error Estimation}
A different vein of explicit model error approximation comes from numerical analysis and control theory: using residuals and adjoint equations to estimate the impact of model errors on quantities of interest (QoIs). These methods often target numerical approximation error (discretization error), but the underlying principles also inform structural error analysis \cite{Roy2011, Oden2013}.

In finite element analysis, \emph{a posteriori} error estimation techniques use the computed solution to evaluate the residual of the differential equations and then solve an adjoint (dual) problem to determine how that residual affects a specific QoI $J(u)$ \cite{Roy2011,SuranaJoy2016}. For a PDE with exact solution $u$ and numerical solution $u_h$, the error in a QoI can be approximated as

\begin{equation}
  J(u) - J(u_h) \;\approx\; \mathcal{R}(u_h, \phi),
  \label{eq:dwr}
\end{equation}

\noindent where $\mathcal{R}$ is a residual functional and $\phi$ is the solution of an adjoint problem linearized around $u_h$ with $J$ as a source term (dual-weighted residual or DWR method). This yields explicit error estimates that can be used to correct $J(u_h)$ (superconvergent estimates) or to drive adaptive mesh refinement strategies \cite{Roy2011}.

Residual-based ideas are also used in model order reduction error estimates. The ROMES method by Drohmann and Carlberg \cite{Drohmann2015} uses inexpensive error indicators (e.g.\ residual norms) as inputs to a statistical model that predicts the true error of a reduced-order model. Manzoni et al. \cite{manzoni2016accurate} extended such approaches to combine reduced-basis a posteriori error bounds with reduction error models, quantifying and mitigating the impact of reduced-order errors on the posterior in Bayesian inverse problems. These techniques explicitly model the reduced-model error using residual information and statistical surrogates.

Adjoint-based sensitivity analysis can also be used to identify which parts of a model contribute most to error in a QoI and to compute optimal model perturbations (bias corrections) in variational DA and optimal control \cite{Tremolet2006, Roy2011}. Although most established in the context of discretization error, residual and adjoint methods provide a rigorous backbone for error assessment and can be combined with statistical discrepancy models for comprehensive UQ \cite{Oden2013, Roy2011}.

\section{Model Error in Data Assimilation, Inverse Problems, and Optimization}
\label{sec:applications}
Many of the techniques discussed in this work, summarized in Table \ref{tab:model-error-approaches}, have been developed with specific application domains in mind. Here we highlight how model error treatment is tailored in three important areas: data assimilation (DA) in geosciences, general inverse problems, and PDE-constrained optimization, including optimal design and control.
From a computational perspective, these approaches are implemented on top of high-dimensional PDE-based forecast models and ensemble or variational algorithms, making model error treatment an integral part of large-scale numerical simulation pipelines.

\begin{table}[h!]
\centering  
\begin{footnotesize}
\renewcommand{\arraystretch}{1.1}
\begin{tabular}{p{0.26\textwidth} p{0.07\textwidth} p{0.60\textwidth}}
\hline
\multicolumn{1}{c}{\textbf{Summary}} &
\multicolumn{1}{c}{\textbf{Type}} &
\multicolumn{1}{c}{\textbf{Description}} \\
\hline
Bayesian approximation error (BAE)
& Implicit
& Treats high–low model discrepancy as a random variable and premarginalizes it, yielding an inflated likelihood covariance. \\

Probabilistic numerics
& Implicit
& Places probability measures on discretization error so that numerical solvers return distributions over solutions. \\

Model error as process noise in DA
& Implicit
& Represents model inadequacy via process noise, stochastic perturbations, covariance inflation, or weak-constraint formulations. \\

Bayesian model discrepancy in calibration
& Explicit
& Adds an explicit discrepancy term to the model output to separate structural error from parameter uncertainty. \\

Gaussian-process discrepancy and multi-fidelity emulation
& Explicit
& Uses Gaussian-process surrogates to emulate and correct discrepancies between fidelities or between model and data. \\

Embedded error models in governing equations
& Explicit
& Introduces additional fields or source terms inside PDE/ODE operators to represent missing or uncertain physics. \\

Machine-learning-based discrepancy correction
& Explicit
& Trains ML models to learn corrective tendencies or operators on top of a physics-based core model. \\

Multi-fidelity and hybrid surrogate modeling
& Explicit
& Combines low- and high-fidelity or data-driven and physics-based models with explicit cross-model error corrections. \\

Residual- and adjoint-based error estimation
& Explicit
& Uses residuals and adjoint solutions to estimate errors in quantities of interest and to drive adaptivity or ROM error models. \\
\hline
\end{tabular}
\end{footnotesize}
\caption{Summary of implicit/explicit strategies for model error treatment considered in this work.}
\label{tab:model-error-approaches}
\end{table}

In \textbf{data assimilation}, operational systems for weather and ocean forecasting have long incorporated model error through process noise, bias terms, and weak-constraint formulations \cite{Tremolet2006, Mitchell2015, Carrassi2018}. 
For example, weak-constraint 4D-Var formulations used in operational and research numerical weather prediction allow bias terms or model-error tendencies in prognostic equations and estimate them alongside the state.
Ensemble-based systems use covariance inflation and stochastic parameterizations to avoid filter divergence and to represent uncertainty from unresolved scales \cite{Mitchell2015, Carrassi2018}. Recent developments combine these ideas with ML-based model error surrogates, as discussed in Section~\ref{sec:DA} and the hybrid modeling section \cite{Farchi2021, Brajard2021}.

In \textbf{inverse problems}, model error is a central concern. The BAE approach allows one to use reduced or simplified forward models while maintaining statistically valid inferences by treating approximation errors as random and premarginalizing them \cite{Kaipio2007, Huttunen2007}. 
In statistical calibration, KOH-type discrepancy models are now standard for calibrating complex computer models to data, with sophisticated treatments of identifiability, experimental design, and multi-fidelity data \cite{Kennedy2001, Arendt2012, Brynjarsdottir2014}. Design of experiments can explicitly target regions that best distinguish parameter uncertainty from model discrepancy, improving inference quality \cite{Arendt2012, Bayarri2007}.

In \textbf{PDE-constrained optimization and control}, model error can lead to suboptimal or even unsafe designs if not accounted for. Multi-fidelity optimization frameworks integrate low- and high-fidelity models, plus error estimators, to ensure that design decisions remain robust when evaluated with more accurate models or reality \cite{Allaire2014, Peherstorfer2018}. Robust optimization and robust control explicitly incorporate model uncertainty -- often represented as bounded perturbations or stochastic discrepancy -- into the optimization criterion, aiming for designs that perform acceptably under plausible model deviations. Residual- and adjoint-based error estimates, together with statistical discrepancy models, provide quantitative information about how model error propagates to design objectives and constraints \cite{Roy2011, Oden2013, Drohmann2015}.

Across these fields, a recurring theme is the need to balance model fidelity, computational cost, and uncertainty. Modern workflows increasingly combine multiple strategies: for example, using BAE to handle reduced-model error in an inverse problem, probabilistic numerics to account for discretization error, and ML-based discrepancy models to capture residual patterns in data assimilation or design loops.


\section{Conclusions and Outlook}
We have surveyed developments in  {treating} model error, distinguishing between implicit approaches (which absorb model inaccuracies into broader uncertainty quantification) and explicit approaches (which seek to represent and correct the errors directly). 
Starting from the basic decomposition, we emphasized that acknowledging model-form and numerical errors is essential for rigorous uncertainty quantification \cite{Roy2011, Oden2013}.

\begin{itemize}
  \item \textbf{Implicit methods} such as the Bayesian approximation error approach enable the use of simplified models with corrected uncertainties by premarginalizing over model discrepancies \cite{Kaipio2007, Huttunen2007, lipponen2018correction, kolehmainen2009approximation}. In data assimilation and forecasting, treating model error as process noise or bias has become standard practice, improving filter stability and forecast reliability \cite{Mitchell2015, Carrassi2018, Tremolet2006}.
  
  \item \textbf{Explicit discrepancy modeling} has evolved to structured, physics-informed discrepancy models \cite{Xiao2016, Zhang2018}, helping to avoid biased parameter estimates and to quantify model inadequacy as a function  {of inputs/conditions/operating regime.} \cite{Brynjarsdottir2014}.
  
  \item \textbf{Probabilistic numerical methods} reframe numerical error as a statistical quantity, integrating discretization uncertainty into scientific conclusions \cite{Hennig2015, Cockayne2019, Conrad2017}. This is increasingly relevant as simulations grow in complexity and scale.
  
  \item \textbf{Machine learning and hybrid approaches} represent a rapidly growing frontier: data-driven correctors trained on model -- truth discrepancies have shown substantial accuracy gains in turbulence modeling, geophysical prediction, and inverse problems \cite{Duraisamy2019, Smyl2021, Brajard2021, Farchi2021, Pathak2018}. Ensuring that these corrections respect physical principles remains an active and critical area of research \cite{Willard2022}.
  
  \item \textbf{Adjoint and residual methods} provide error bounds and estimators for QoIs, particularly for discretization and reduced-model error \cite{Roy2011, Drohmann2015}. They complement statistical methods and can be combined with them for comprehensive UQ \cite{Oden2013}.
\end{itemize}

\noindent Looking ahead, important directions are anticipated to include:

\begin{itemize}
  \item \textbf{Automated model error discovery and diagnosis}, using ensembles of models and ML to localize deficiencies and suggest missing physics.
  
  \item \textbf{Standards for UQ with model error}, especially in safety-critical fields where design factors of safety could be better informed by quantitative discrepancy estimates rather than purely heuristic margins.
  
  \item \textbf{Integration of multi-source and multi-fidelity data}, constraining discrepancy models with heterogeneous observational streams and model hierarchies \cite{Peherstorfer2018, FernandezGodino2023}.
  
  \item \textbf{Real-time model error correction} within digital twin frameworks, requiring robust, online estimation of discrepancy under computational constraints.
  
  \item \textbf{Extrapolation-aware modeling}, developing methods that reliably signal or inflate uncertainty when predictions venture beyond the domain in which discrepancy models were calibrated.
  
  \item \textbf{Generative diffusion models}  {are emerging and may become useful} for high dimensional discrepancy learning \cite{chen2024opportunities,shu2023physics}, providing expressive non-Gaussian representations of structural, numerical, and subgrid model errors. 
  These approaches may amplify ML-based corrector strategies (including {PINN} approaches) with probabilistic numerics and Bayesian discrepancy modeling, enabling scalable uncertainty quantification for PDE-based simulations.
\end{itemize}

Ultimately,  {the credibility of simulation and data-driven prediction hinges on transparent, defensible representations of model error and its impact on uncertainty and bias}.
The diverse set of tools reviewed here -- from BAE and probabilistic numerics to GP-based discrepancy models, embedded error representations, ML correctors, and adjoint-based error estimators -- provide a rich toolbox for practitioners. Combining these methods thoughtfully can yield models that are not only more accurate but also more honest about their limitations, enabling more trustworthy computational science and engineering.
We  {anticipate} that many of the strategies surveyed here will increasingly be integrated into mainstream computational physics workflows, informing the design of numerical methods and simulation codes that are not only more accurate but also quantitatively aware of their own limitations.

\section*{Acknowledgment} 
DS would like to acknowledge the School of Civil and Environmental Engineering at Georgia Institute of Technology for supporting this work.

\section*{Funding Declaration} 
No funding was provided for this work.

\section*{Author contributions}
DS: Conceptualization; Methodology; Validation; Formal analysis; Investigation; Data curation; Writing – original draft; Writing – review \& editing; Visualization; Project administration

\bibliographystyle{unsrt}
\bibliography{refs}
\end{document}